\documentclass[twocolumn]{aastex62}

\defcitealias{finley2018dipquadoct}{FM18}
\usepackage{mhchem}

\usepackage{array,graphicx}
\usepackage{booktabs}
\usepackage{pifont}

\graphicspath{{./}{figures/}}

\received{--}
\revised{--}
\accepted{--}

\shorttitle{Solar Angular Momentum-loss with \textit{PSP}}
\shortauthors{Finley et al.}

\begin{document}

\title{The Solar Wind Angular Momentum Flux as Observed by \textit{Parker Solar Probe}}

\correspondingauthor{Adam J. Finley}
\email{*af472@exeter.ac.uk}

\newcommand{\exeter}{\affiliation{University of Exeter, Exeter, Devon, EX4 4QL, UK}}
\newcommand{\cea}{\affiliation{Laboratoire AIM, DRF/IRFU/D\'epartement d'Astrophysique, CEA-Saclay, 91191, Gif-sur-Yvette, France}}
\newcommand{\reading}{\affiliation{University of Reading, Reading, Berkshire, RG6 6BB, UK}}
\newcommand{\irap}{\affiliation{IRAP, Universit\'e Toulouse III - Paul Sabatier, CNRS, CNES, Toulouse, France}}
\newcommand{\sao}{\affiliation{Smithsonian Astrophysical Observatory, Cambridge, MA, USA}}
\newcommand{\umich}{\affiliation{University of Michigan, Ann Arbor, MI, USA}}
\newcommand{\ucb}{\affiliation{University of California, Berkeley, CA, USA}}

\author[0000-0002-3020-9409]{Adam J. Finley}\exeter
\author[0000-0001-9590-2274]{Sean P. Matt}\exeter
\author[0000-0002-2916-3837]{Victor R\'eville}\irap
\author[0000-0001-8247-7168]{Rui F. Pinto}\cea\irap
\author[0000-0003-2061-2453]{Mathew Owens}\reading

\author[0000-0002-7077-930X]{Justin C. Kasper}\umich\sao
\author[0000-0001-6095-2490]{Kelly E. Korreck}\sao
\author[0000-0002-3520-4041]{A. W. Case}\sao
\author[0000-0002-7728-0085]{Michael L. Stevens}\sao
\author[0000-0002-7287-5098]{Phyllis Whittlesey}\ucb
\author[0000-0001-5030-6030]{Davin Larson}\ucb
\author[0000-0002-0396-0547]{Roberto Livi}\ucb




\begin{abstract}
The long-term evolution of the Sun's rotation period cannot be directly observed, and is instead inferred from trends in the measured rotation periods of other Sun-like stars. Assuming the Sun spins-down as it ages, following \textit{rotation rate} $\propto$ \textit{age}$^{-1/2}$, requires the current solar angular momentum-loss rate to be around $6\times 10^{30}$erg. Magnetohydrodynamic models, and previous observations of the solar wind (from the \textit{Helios} and \textit{Wind} spacecraft), generally predict a values closer to $1\times 10^{30}$erg or $3\times 10^{30}$erg, respectively. Recently, the \textit{Parker Solar Probe} (PSP) observed tangential solar wind speeds as high as $\sim50$km/s in a localized region of the inner heliosphere. If such rotational flows were prevalent throughout the corona, it would imply that the solar wind angular momentum-loss rate is an order of magnitude larger than all of those previous estimations.  In this letter, we evaluate the angular momentum flux in the solar wind, using data from the first two orbits of PSP.  The solar wind is observed to contain both large positive (as seen during perihelion), and negative angular momentum fluxes.  {We analyse two solar wind streams that were repeatedly traversed by PSP; the first is a slow wind stream whose average angular momentum flux fluctuates between positive to negative, and the second is an intermediate speed stream containing a positive angular momentum flux (more consistent with a constant flow of angular momentum).} When the data from PSP is evaluated holistically, the average equatorial angular momentum flux implies a global angular momentum-loss rate of around $2.6-4.2\times 10^{30}$ erg (which is more consistent with observations from previous spacecraft).
\end{abstract}



\section{Introduction}
The solar wind is steadily removing angular momentum (AM) from the Sun \citep{weber1967angular, mestel1968magnetic}. This can be measured in-situ by evaluating the mechanical AM flux in the solar wind particles, and the stresses in the interplanetary magnetic field \citep{lazarus1971observation, pizzo1983determination, marsch1984distribution, li1999magnetic, finley2019direct}.
The value of the current solar AM-loss rate is a useful test of models which attempt to describe the rotation-evolution of low-mass stars \citep[i.e. $M_*\leq 1.3M_{\odot}$; e.g.][]{gallet2013improved, gallet2015improved, brown2014metastable, johnstone2015stellar, matt2015mass, amard2016rotating, amard2019first, blackman2016minimalist, sadeghi2017semi, garraffo2018revolution, see2018open}.
Such stars have magnetic activity which is directly linked to their rotation rates \citep{wright2011stellar, wright2016solar}.
One consequence is that the habitability of exoplanets will likely depend somewhat on the rotation rate of their host star, and how it has varied in the past \citep[e.g.][]{johnstone2015evolution, gallet2017impacts}.

\begin{table*}
\caption{Observed Solar Wind Angular Momentum Fluxes}
\label{AMvalues}
\center
\setlength{\tabcolsep}{2pt}
  \begin{tabular}{c|cc|c|c|c}
      \hline\hline
Spacecraft & Component &$\langle r^2 F_{AM}\rangle$ & Protons/ & Radial	&	Source	\\
 Name && [$\times 10^{30}$erg/ster]  & Magnetic	& Distance [au] & 			\\	\hline
\textit{Parker Solar Probe*} & {total} & {0.31(0.50)} & 0.9(3.2) & 0.16-0.7 &  This Work  \\
 & \color{red} protons & \color{red} 0.15(0.38)& &  &    \\
 & alphas & -& &  &    \\
 & \color{blue} magnetic & \color{blue} 0.16(0.12)& &  &    \\ \hline
\textit{Wind} & {total} & {0.39} & 2.4&  1 & \begin{NoHyper}\cite{finley2019direct}\end{NoHyper}  \\
 & \color{red} protons & \color{red} 0.29 & &  &    \\
 & alphas & -0.02 & &  &    \\
 & \color{blue} magnetic & \color{blue} 0.12 & &  &    \\ \hline
\textit{Helios} & {total} & {0.20} & 1.1& 0.3-1 &   \begin{NoHyper}\cite{pizzo1983determination}\end{NoHyper}  \\
 & \color{red} protons & \color{red} 0.17 & &  &    \\
 & alphas & -0.13 & &  &    \\
 & \color{blue} magnetic & \color{blue} 0.15 & &  &    \\ \hline
\textit{Mariner 5} & {total} & {$\bf \sim 1.2$}& $\sim 4.3$& 0.6 & \begin{NoHyper}\cite{lazarus1971observation}\end{NoHyper}  \\
 & \color{red} protons & \color{red} $\sim1$ & &  &    \\
 & alphas & -& &  &    \\
 & \color{blue} magnetic & \color{blue} 0.23 & &  &    \\ \hline \hline
  \end{tabular}

      \small
      \item *Values for PSP are the averaged values from Figure \ref{latitudinallyAveraged} in the format E01(E02).

\end{table*}

Stellar convection, rotation, and magnetic field generation are all intricately linked for low-mass stars \citep[see review of][]{brun2017magnetism}.
In general, this causes the rotation rates of Sun-like stars on the main sequence to broadly follow an approximate relationship where \textit{rotation rate} $\propto$ \textit{age}$^{-1/p}$ \citep{skumanich1972time, soderblom1983rotational, barnes2003rotational, barnes2010angular, lorenzo2019constraining}, where $p$ is observed to be around 2.
However, with the increasing number of rotation period observations \citep{agueros2011factory, agueros2017setting, mcquillan2013measuring, nunez2015linking, rebull2016rotation, covey2016rapidly, douglas2017poking, nascimento2020rotation}, it has become clear that rotation may not always follow a simple, single power-law in time \citep[e.g.][]{davenport2018rotating, metcalfe2019understanding, reinhold2019transition}.  For example, there may be a temporary ``stalling'' of the spin-down during the early main sequence \citep{curtis2019temporary} or a significant reduction of spin-down torques (or equivalently a larger value of $p$) at approximately the solar age \citep[][]{van2016weakened}.

Previous measurements of the solar wind AM flux using the \textit{Helios} \citep{pizzo1983determination, marsch1984distribution}, and \textit{Wind} \citep{finley2019direct} spacecraft, have suffered from pointing errors and/or large uncertainties in the detection of the tangential solar wind speed at 1au (with an amplitude around 1-5 km/s), but these studies suggest an average equatorial {flow of AM per solid angle} $\langle r^2 F_{AM}\rangle$ of around $0.3\pm 0.1\times 10^{30}$erg/steradian.
The role of the solar wind alpha particles in carrying AM is unclear. Measurements from the \textit{Helios} spacecraft indicate they carry a net negative AM flux, whereas measurements from \textit{Wind} suggest they have a negligible contribution to the total AM flux. Notable values from previous works can be found in Table \ref{AMvalues}.
A common factor in most previous observations, including those that use older spacecraft like \textit{Mariner 5} \citep[][]{lazarus1971observation}, is the presence of localized wind streams that can carry a net negative AM flux. A likely mechanism to generate these negative streams is wind stream-interactions, i.e. when fast wind streams catch up to, and collide with, slow wind streams. It is expected that the fast wind is then significantly deflected, given its lower density, in the direction opposite of rotation. Therefore the fast solar wind should carry the majority of the observed negative AM flux \citep[which is shown in][]{finley2019direct}.

\begin{figure*}
   \begin{center}
    \includegraphics[trim=2.cm 4.cm 0.5cm 2.cm, clip, width=\textwidth]{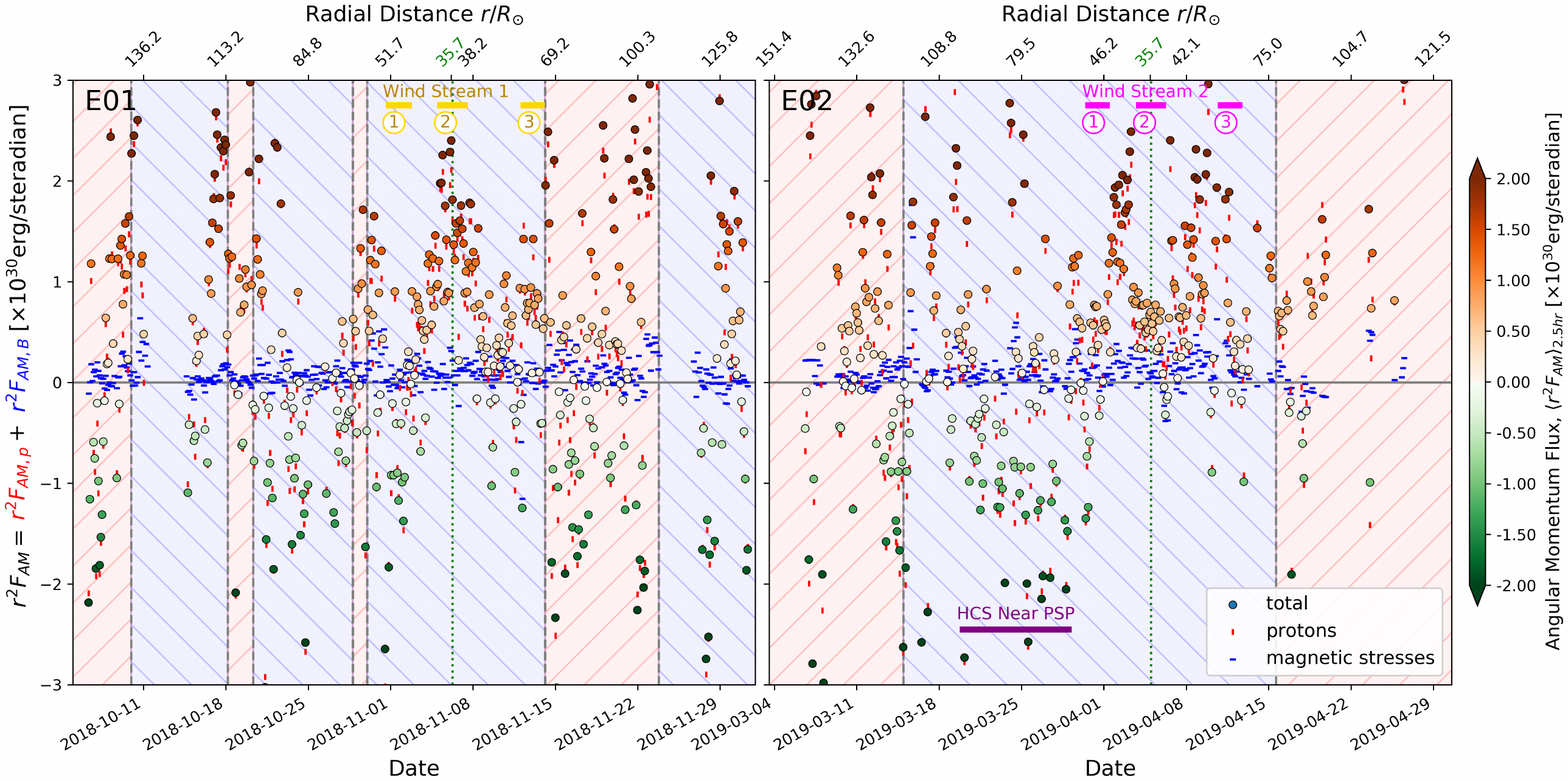}
  \end{center}
   \caption{Solar wind AM flux observed by PSP, in the protons (red vertical ticks), magnetic field stresses (blue horizontal ticks), and their sum (coloured circles), averaged over 2.5 hour intervals versus time. The left panel shows the first encounter E01 (5th Oct - 2nd Dec, 2018), and the right panel shows the second encounter E02 (3rd Mar - 30th Apr, 2019). The perihelia of each orbit are indicated by green dotted lines. The HCS crossings from Figure \ref{trajectory} are indicated with grey dashed lines, with the background color/hatching corresponding to the global magnetic field polarity. A sustained period of negative AM flux, which coincides with PSP being in close proximity to the HCS is indicated in purple, and is also highlighted in Figure \ref{trajectory}. {We identify repeated crossings of a slow solar wind stream during E01 (in yellow), and an intermediate speed stream during E02 (magenta). These wind streams are shown in more detail in Figures \ref{trajectory} and \ref{windStreams}.}}
   \label{fluxes}
\end{figure*}

{During its first two orbits, \textit{Parker Solar Probe} (PSP) observed} the solar wind close to the Sun ($\sim 36R_{\odot}$) to have tangential speeds of up to $\sim 50$km/s \citep{kasper2019alfvenic}, which is far greater than the expected $1-5$km/s from previous \cite{weber1967angular} wind modelling \citep[e.g.][]{reville2020role}. At face value, this implies a larger AM-loss rate from the Sun than previously thought \citep[see also estimates in][]{finley2018effect, finley2019solar}.
However, in this letter we argue that the observations made by PSP during its perihelion passes are not necessarily representative of the global AM-loss rate. As such, by taking into account the spatial variation of the AM flux, our average values from PSP are closer to previous spacecraft observations.

\section{Data}
In this work we utilise the publicly available data from both the Solar Wind Electrons, Alphas and Protons (SWEAP) instrument suite\footnote{http://sweap.cfa.harvard.edu/pub/data/sci/sweap/spc/L3/ - Accessed March 2020.} \citep{kasper2016solar}, and the FIELDS instrument suite\footnote{http://research.ssl.berkeley.edu/data/psp/data/sci/fields/l2/ - Accessed March 2020.} \citep{bale2016fields}, during the first two orbits of PSP. The Solar Probe Cup (SPC) \citep{case2020solar}, part of the SWEAP instrument suit, is capable of measuring the velocity distribution and density of the solar wind particles using moment-fitting algorithms which return the bulk characteristics of the particle populations. SPC operates at a varying data cadence during the orbit of PSP around the Sun, with its highest sampling rate inside 0.25au. Vector magnetic field data is collected by the FIELDS instrument suite at various time resolutions \citep[e.g.][]{bale2019highly}. For this work, we use the minute cadence data and interpolate this down to the variable time resolution of the SPC data.

\begin{figure*}
   \begin{center}
    \includegraphics[trim=3.cm 1.cm 2.cm 0.cm, clip, width=\textwidth]{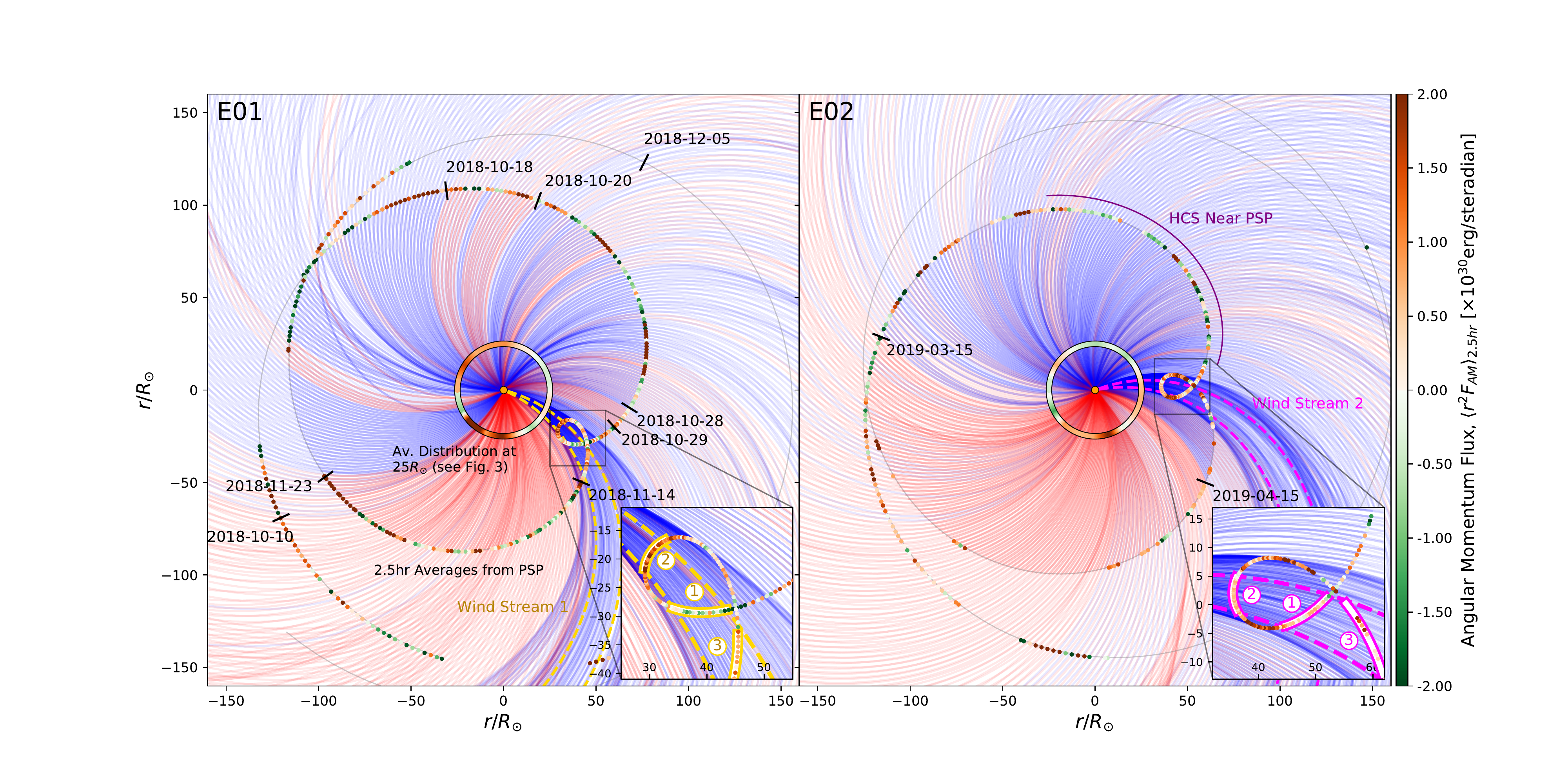}
  \end{center}
   \caption{The trajectory of PSP (grey line) in a reference frame co-rotating with the Sun, projected onto the equatorial plane (as viewed from above the North pole), with the Sun at the center. The first encounter E01 (perihelion 6th November 2018) is in the left panel, and the the second encounter E02 (perihelion 4th April 2019) is in the right panel. The AM flux in the solar wind as observed by PSP (in the protons plus magnetic field stresses, as in Figure \ref{fluxes}), is then shown using coloured circles that each represent the 2.5 hour average values. Using the radial wind speed observed by SPC, the connectivity of the magnetic field in the inner heliosphere is visualised with Parker spiral magnetic field lines, which are initialised along PSP's trajectory at 2.5 hour increments (only when $r < 124R_{\odot}$). Each field line is coloured by the magnetic field polarity observed by FIELDS, averaged over the 2.5 hour increment (red is positive, blue is negative). Significant reversals in the observed magnetic field polarity, likely caused by crossing the HCS, are indicated with black lines along PSP's trajectory with their associated dates. {Times when PSP crossed the same solar wind stream are highlighted in the inset figures (yellow for E01, and magenta for E02), with each crossing labelled a number in the order that they were encountered. Dashed lines show the expected boundaries of each wind stream based on parker spiral trajectories that use the average radial wind speed from each wind stream.} The inner annulus at $25R_{\odot}$ displays the average AM flux from Figure \ref{longitudinallyAveraged} where the data are ballistically mapped to $25R_{\odot}$ using Parker spiral trajectories and then binned by Carrington longitude. }
   \label{trajectory}
\end{figure*}

During PSP's first orbit we use data from the 5th October 2018 to the 2nd December 2018, with perihelion occurring on the 6th November. This interval is henceforth referred to as E01.
Data for the second orbit is available from the 3rd March 2019 to the 30th April 2019, with perihelion occurring on the 4th April; similarly this period is referred to as E02. For the first orbit, we supplement the public data during the inbound phase (during October 2018 only), with data supplied by the instrument team ({\color{blue}SWEAP team, private communication}). {During the first two orbits of PSP the alpha particle moments were not well recovered, and so in this letter we focus on the proton observations.} We remove proton and magnetic field data that have been flagged by the instrument teams as containing bad/problematic values. We also evaluate the data taken during the third orbit of PSP (E03), though this dataset is incomplete due to a technical failure of the SPC instrument on approach to perihelion. Therefore in this letter we focus on the first two orbits and present the third orbit as supplemental information in Appendix \ref{E03}.

\section{Observed Solar Wind Angular Momentum Flux}

\begin{figure*}
   \begin{center}
    \includegraphics[trim=.cm .cm .cm 0.cm, clip, width=0.9\textwidth]{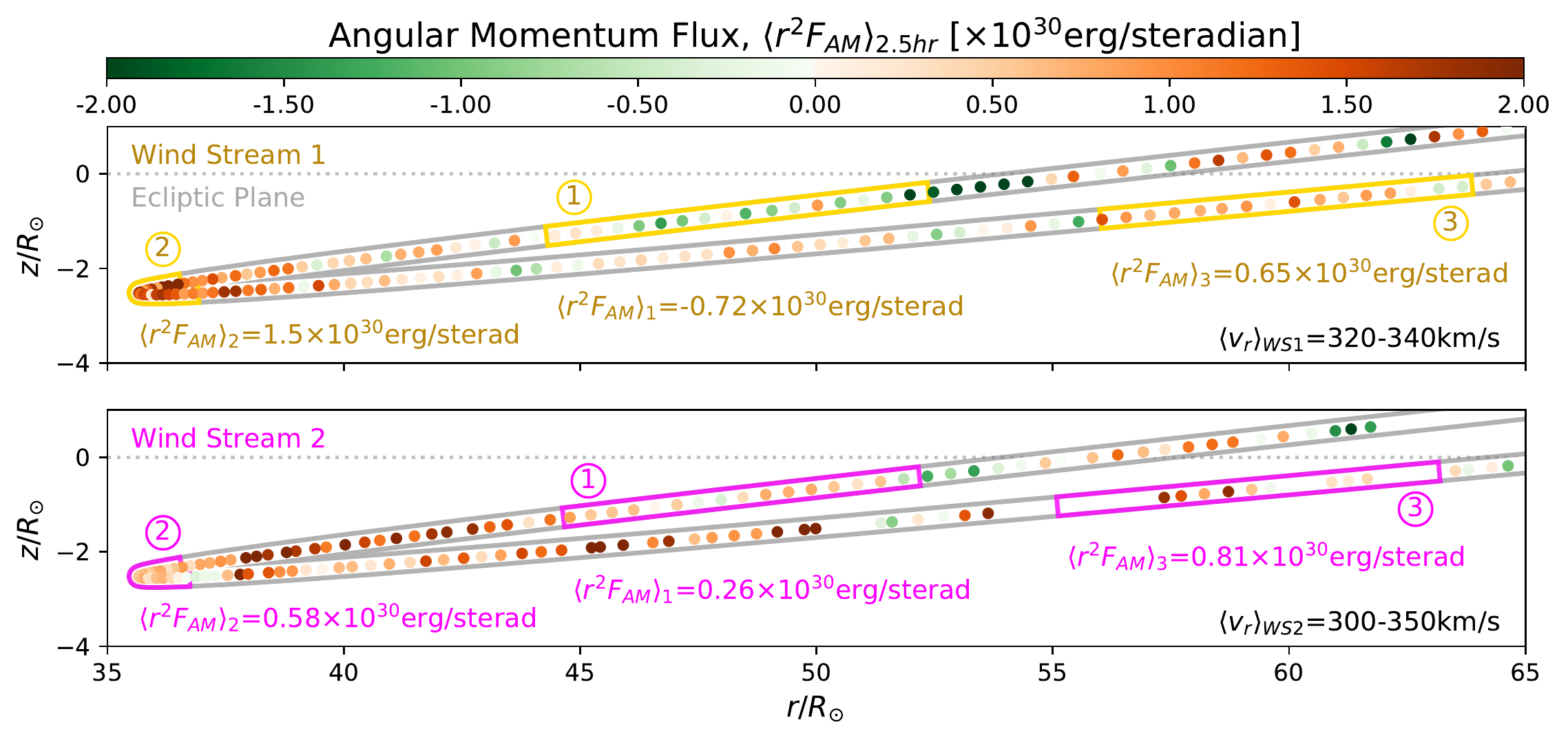}
  \end{center}
   \caption{{The trajectory of PSP plotted as a function of height from the ecliptic plane and cylindrical radius, shown in grey for E01 (top) and E02 (bottom). The 2.5 hour average AM flux is shown with coloured circles, and the times that PSP crossed Wind Stream 1 during E01, and Wind Stream 2 during E02, are highlighted in yellow and magenta respectively. The average AM flux during each crossing is also displayed.}}
   \label{windStreams}
\end{figure*}

Using the observations from PSP, we evaluate the solar wind AM flux ($F_{AM}$) as a sum of the mechanical AM carried by the protons ($F_{AM,p}$) and the transfer of AM through magnetic field stresses ($F_{AM,B}$), at the cadence of the SPC instrument. This is given by,
\begin{align}
r^2F_{AM}=& r^2F_{AM,p} + r^2F_{AM,B},\nonumber\\
=& r^3\sin\theta\rho v_r v_t -r^3\sin\theta\frac{B_t B_r}{4\pi},
\label{AMequation}
\end{align}
where $r$ is the radial distance of PSP, $\theta$ is its colatitude, $\rho$ is the proton density, $v_r$ is the radial wind speed of the protons, $v_t$ is the tangential wind speed of the protons, $B_r$ is the radial magnetic field strength, and $B_t$ is the tangential magnetic field strength. Here a factor of $r^2$ has been included to remove the dependence of the AM flux on radial distance, as it is only the poloidal vorticity-current stream function (i.e. $r\sin\theta v_t-r\sin\theta B_t B_r/4\pi\rho v_r$) that is a conserved quantity along magnetic field lines under the assumptions of ideal magnetohydrodynamics \citep[see][for the correct nomenclature]{goedbloed2019magnetohydrodynamics}.
{The quantity presented throughout this letter $r^2F_{AM}$ is the flow of AM per solid angle (due to this normalisation with radius), though we often refer to this as an AM flux for simplicity.}
This is the same quantity evaluated by previous authors using other spacecraft, and so can be directly compared (see Table \ref{AMvalues}). However, it is worth noting that equation (\ref{AMequation}) does not include the effect of thermal pressure anisotropies which could influence the magnetic stress term \cite[as discussed in][]{reville2020role}.  When summing over all longitudes, equation (\ref{AMequation}) effectively assumes that any AM flux due to thermal pressure anisotropies will sum to zero.
Observations suggest the solar wind between 0.3-1au has a mostly isotropic plasma pressure \citep[e.g.][]{marsch1984helios}.

Due to small scale fluctuations in the solar wind, it is necessary to average the AM flux on a sufficient spatial and/or temporal scale to recover the character of the large-scale solar wind flow. {It is of course possible that these fluctuations transport an additional AM flux to that of the bulk solar wind, for example via compressible MHD waves \citep[e.g.][]{marsch1986acceleration}. However, at present we focus on constraining the properties of the bulk solar wind, as this is likely where the majority of the AM flux is contained.} In Figure \ref{fluxes}, we present the {flow of AM per solid angle ($r^2F_{AM}$)} averaged over 2.5 hour intervals versus time for both E01 and E02. The proton AM flux is shown with red vertical ticks, the magnetic field stresses with blue horizontal ticks, and their total with coloured circles. {The signal to noise on the SPC tangential wind speed observations generally decreases with radial distance, and so the percentage uncertainty increases. However the proton AM flux varies on a scale that is generally larger than these uncertainties.}

These observations, indicate that the solar wind AM flux has a substantial spatial variation.
PSP even observes significant periods of time with a sustained net negative AM flux (negative flux implies the addition of AM to the Sun). The most obvious example occurred when PSP was in close proximity to the Heliospheric Current Sheet (HCS), which is annotated in purple in the right panel of Figure \ref{fluxes}. In contrast to the protons, the magnetic field stresses do not show much variability (on average around $\sim 0.1 \times 10^{30}$erg/steradian) and are comparable in strength to previous spacecraft observations. Additionally, the uncertainties on a given measurement of $F_{AM,B}$ are much lower than $F_{AM,p}$ because the magnetic field direction is generally not as radial as the solar wind velocity. {Note that our averaging timescale is much greater than that of fluctuations due to ``switchbacks'' \citep[e.g.][]{mcmanus2020cross}, and so the magnetic stress term here relates to large-scale deviation of the interplanetary magnetic field from the radial direction. Future works could consider the effect of these switchbacks on the amount of AM stored in the magnetic field. However, given the observed structure of these fluctuations \citep[switchback rotation angles are investigated in][]{mozer2020switchbacks}, the momentum imparted to the plasma during relaxation of the magnetic field is likely directed radially on average.}

Figure \ref{trajectory} shows the same 2.5 hour averages of the {flow of AM per solid angle}, now along the trajectory of PSP during E01 and E02 (with coloured circles). In the background of each panel, the polarity of the interplanetary magnetic field during each 2.5 hour interval is extrapolated into a Parker spiral using the proton radial wind speed as measured by SPC. This helps to visualise the large-scale structure of the magnetic field in the inner heliosphere. Significant magnetic field polarity reversals are highlighted in both Figures \ref{fluxes} and \ref{trajectory}.
The variation of the AM flux during the closest approaches of PSP are most clear in the zoomed insets at the bottom right corner of each panel.
During PSP's first perihelion, the proton AM flux increases with decreasing radial distance to the Sun, whereas during its second perihelion the proton AM flux is largest for the inbound and outbound observations and is decreased during closest approach (this dip coincides with a sharp decrease in the proton mass flux).
Although there are differences between the AM flux during both perihelia (perihelia are also identified with green dotted lines in Figure \ref{fluxes}), the large-scale organisation of the AM flux, i.e. the locations of the strong positive/negative AM fluxes, show similarities between E01 and E02.

{During the closest approach of PSP in E01, the same slow solar wind stream, hereafter referred to as ``Wind Stream 1'', was crossed three times (highlighted in yellow). Similarly, during E02 PSP crosses another designated ``Wind Stream 2'', though the evidence for this being the same stream throughout PSP's observations is less convincing. {The likely origins of these wind streams are discussed in detail within \cite{Panasenco2020ApJS}.} In Figure \ref{windStreams} we show the latitudinal extent of PSP's orbit during the first two perihelia and highlight these two wind steams. For each crossing we compute the average flow of AM per solid angle measured by PSP during the intervals. Note these crossing are a few days apart. Theoretically, for the same solar wind stream the AM flux is expected to be constant (apart from during interactions with other wind streams at stream interfaces). However within Wind Stream 1, the AM flux in the solar wind can be seen to vary from positive to negative. This is a surprising result as Wind Stream 1 is the best constrained stream, and so this can only be explained in a few ways: 1) AM flux varies substantially in space through individual wind streams, and perhaps these fluctuations relate to PSP's location with respect to the boundaries of other wind streams, 2) AM flux varies substantially in time, and perhaps these fluctuations relate to the formation of the slow solar wind \citep[see][]{rouillard2020relating, reville2020tearing}, or 3) the protons and magnetic field stresses do not account for the total AM flux, i.e. the alpha particles, pressure anisotropies, or other processes, are required to explain the observations. In comparison to Wind Stream 1, Wind Stream 2 appears to be closer to having a steady AM flux from crossing to crossing, despite the stream likely containing wind from various sources. During the perihelion of E02, there is a notable difference in the AM flux carried by Wind Stream 2 (an intermediate speed wind stream) and the slow wind streams either side of it. In this case the faster wind contains roughly a quarter of the AM flux observed in the slow wind either side. This observation, along with our analysis of the two wind streams may suggest that intermediate/fast solar wind streams contain a smaller but less time-varying AM flux, than the slow wind streams which host a larger, temporally and/or spatially varying, AM flux.}

\begin{figure*}
 \centering
  \includegraphics[trim=2.5cm 1cm 2.5cm 0.5cm,clip, width=\textwidth]{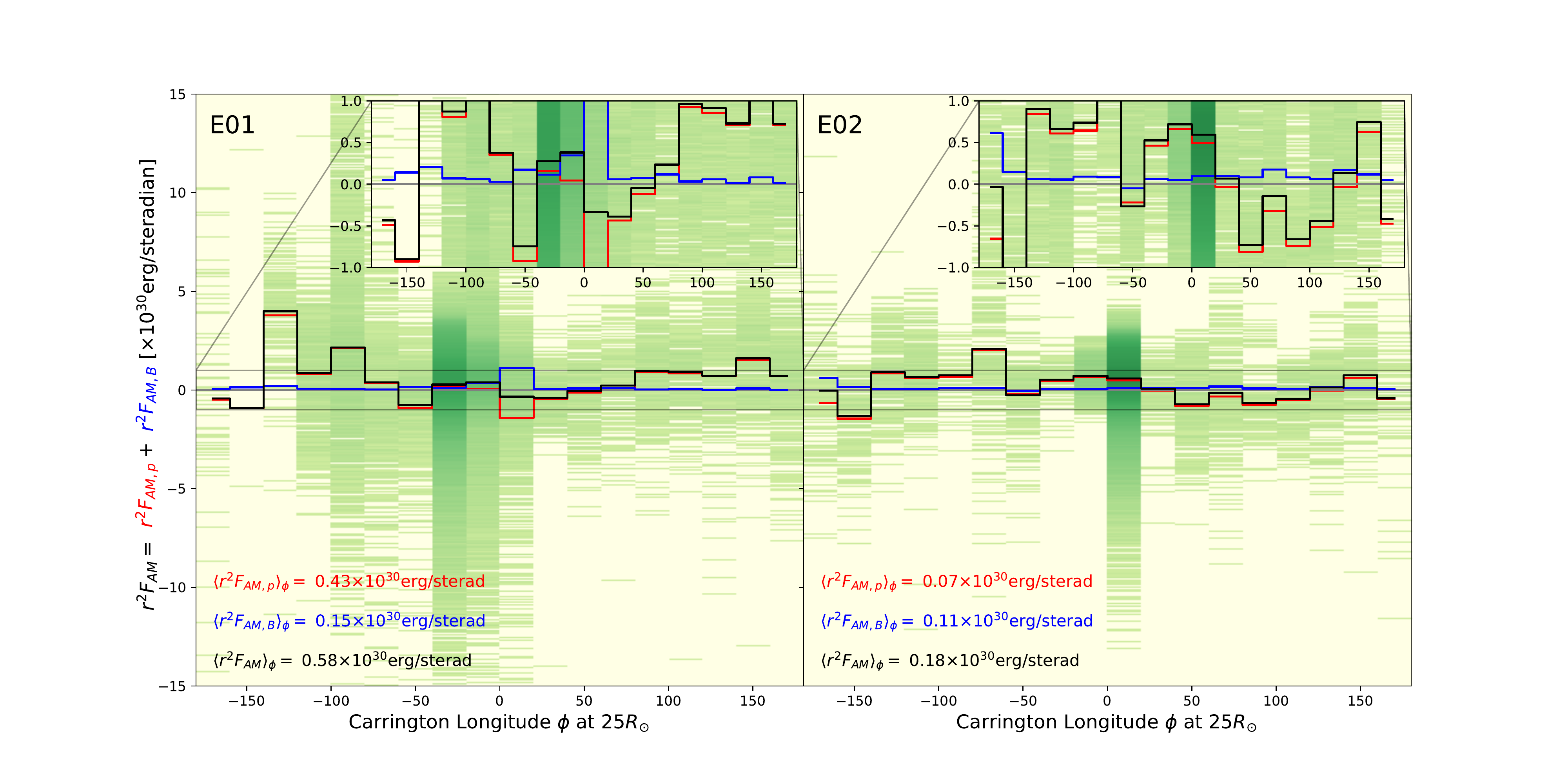}
   \caption{2D histogram of solar wind AM flux (in the protons and magnetic field stresses), versus Carrington longitude, for both E01 (left) and E02 (right). The observations have been ballistically mapped to $25R_{\odot}$ using Parker spiral trajectories with their observed radial wind speeds from SPC, and binned in  $20^{\circ}$ bins. Darker green shades show an increased frequency of observation in the 2D histogram. Note that SPC sampled the solar wind at a much higher cadence during perihelion, which is clearly visible. The average AM flux for each bin as a function of Carrington longitude is then over plotted with a black line. The average values considering just the protons (red), and magnetic field stresses (blue), are also shown. Finally, the sum over all Carrington longitude bins is given in the bottom left text.}
   \label{longitudinallyAveraged}
\end{figure*}

\begin{figure*}
 \centering
  \includegraphics[trim=2.5cm 1cm 2.5cm 0.5cm,clip, width=\textwidth]{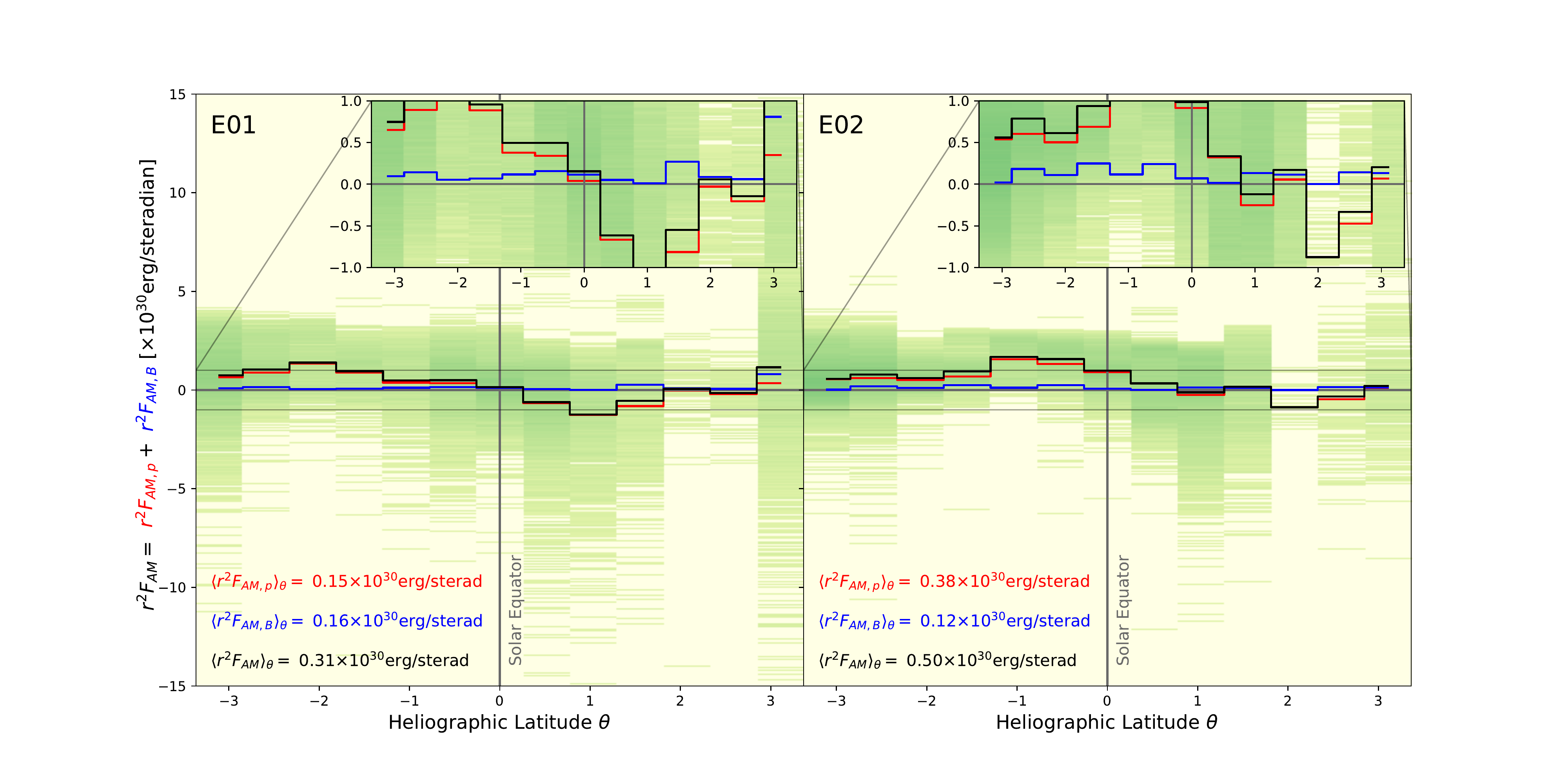}
   \caption{AM flux versus heliographic latitude, in the same format as Figure \ref{longitudinallyAveraged}.
   The average AM flux for each $0.5^{\circ}$ bin as a function of latitude is shown for the protons (red), magnetic field stresses (blue), and their total (black).}
   \label{latitudinallyAveraged}
\end{figure*}

Given the variability of the AM flux, it is difficult to disentangle structure in the solar wind due to PSP's varying radial distance, heliographic latitude, and Carrington longitude.
As an alternative to the temporal averages shown in Figure \ref{fluxes}, Figures \ref{longitudinallyAveraged} and \ref{latitudinallyAveraged} show the AM fluxes binned in Carrington longitude and heliographic latitude, respectively.
Within each spatial bin, the frequency of observing a given value of AM flux colours each 2D histogram (with darker green being higher frequency). The average value from each bin is highlighted in black, along with the averages for just the proton component (red), and just the magnetic stresses (blue).
When the data are binned by Carrington longitude (Figure \ \ref{longitudinallyAveraged}), both E01 and E02 show the AM flux to vary from positive to negative values (this is clearer in the upper right inset figures), with the source of the variation being the proton AM flux (as with the temporal averaging). The AM flux averaged in this way is also shown in an annulus in Figure \ref{trajectory} for both E01 and E02, for visual comparison.
When the data are binned by heliographic latitude, the AM flux is observed to have a clearer structure, shown in Figure \ref{latitudinallyAveraged}.
The inset figures show an approximately sinusoidal variation for both E01 and E02, but with the latitude dividing the positive and negative wind streams seemingly shifted. Again the magnetic stresses do not show much variability with latitude, in comparison to the proton AM flux.

It remains unclear which binning technique best represents the average equatorial AM flux ($F_{AM,eq}$) in the solar wind, and the range of values likely represents the systematic uncertainties arising from the choice of binning method. The highest and lowest values ($r^2F_{AM,eq}=0.18\times 10^{30}$erg/steradian and $0.58\times 10^{30}$erg/steradian) are found by binning the data by Carrington longitude for E01 and E02 respectively.
Hereafter, we adopt the average values when binning the data in latitude, due to the similarity in outcome for both orbits, and these values are given in the top entries in Table \ref{AMvalues}.
For a \cite{weber1967angular} wind, mechanical AM is gained by the particle population from the stresses in the magnetic field as the wind travels through the heliosphere. This means the value of $F_{AM,p}/F_{AM,B}$ for a solar wind parcel should increase with radial distance. The precise value of this ratio and how it varies also provides information about the Alfv\'en point \citep{marsch1984distribution}. This ratio and the radial distance of each spacecraft from previous calculations are shown in Table \ref{AMvalues}. However given the variability of the AM flux with solar cycle \citep[see][]{finley2018effect}, and the varying precision of each instrument, it is difficult to draw any conclusions from this.

The cause(s) of the spatially varying AM flux are unknown, {and there does not appear to be a simple correlation between the enhanced tangential wind speeds and the presence of switchbacks in the solar wind \citep{kasper2019alfvenic}.}
Previously, wind stream-interactions have been used to explain the observed variation in $v_t$ \citep[e.g.][]{pizzo1978three}, though the values observed by PSP are far larger than magnetohydrodynamic models would predict.
For example, the modelling of E01 by \cite{reville2020role} found a similar trend in positive and negative $v_t$, due to stream-interactions, but with an amplitude of only $\pm 5$km/s. For this reason \cite{reville2020role} suggested that pressure anisotropies, which modify the balance of magnetic field stresses and thus how much AM they transfer to the solar wind particles, might explain some (but not all) of the observed trends in $v_t$ with radial distance. Another potential source of the spatially varying AM flux is from magnetic field foot point motions in the photosphere, caused by circulation of the open magnetic flux \citep[][]{crooker2010suprathermal, fisk2020global}. Additionally, the coherent structure in the AM flux between orbits may also indicate a connection with the Sun's large-scale magnetic field. For example the HCS was likely similar in shape between E01 and E02 and so could have played a role in organising the AM flux.

In our analysis, we have used the data as is, and considered the spatial variation of the AM flux and its conflation with the trajectory of PSP. However it is important to acknowledge the uncertainties on these measurements, especially the measurements of $v_t$ from SPC, which have yet to be fully explored. {For example, during more recent encounters in which the Solar Probe Analysers \citep[SPAN,][]{whittlesey2020solar} have been able to measure the proton velocities concurrently with SPC, there are discrepancies which have yet to be resolved.} Over the course of PSP's mission lifetime, as the instrument characteristics are better determined, our understanding of the relative contribution of physical flows and pointing error will increase. It is expected that the signal to noise of SPC, and PSP in general, is significantly higher than previous spacecraft, such as \textit{Helios} and \textit{Wind} \citep[see][for further details about SPC]{case2020solar}.


\section{Conclusion}

We have shown that PSP observed significant spatial variability in solar wind AM flux, with some coherent features between the first two orbits.
In both orbits we find wind streams that carry positive and negative AM fluxes which are separated in longitude and latitude.
{We evaluate two different winds streams which are repeatedly crossed by PSP around each perihelion. From this analysis we show that the AM flux within a given stream can vary substantially, with the slow wind stream (observed during E01) having the largest variations (from positive to negative values), and the intermediate wind stream (observed during E02) being closer to a steady-state flow. This contrast may be introduced by their different solar wind origins, however at present there are not enough observations to constrain this.}
By averaging the data holistically we are able to produce smaller values for the equatorial AM flux than would be inferred by using only data from the closest approaches of PSP.
These values are much closer to previous measurements from a variety of spacecraft at larger radial distances (see Table \ref{AMvalues}), where observations had previously been averaged over $\sim 27$ day intervals to improve the signal to noise \citep[e.g.][]{finley2019direct}.

Assuming the solar wind AM flux is, on the large-scale, distributed as $F_{AM}(\theta)\approx F_{AM,eq}\sin^2\theta$, the global AM-loss rate implied by the average PSP observations ($r^2F_{AM,eq}=0.31-0.50\times 10^{30}$erg/steradian) is $2.6-4.2\times 10^{30}$erg. This value is around a factor of 2 smaller than what would be expected from a \cite{skumanich1972time} rotation period evolution \citep[\textit{rotation rate} $\propto$ \textit{age}$^{-1/2}$; e.g.][]{matt2015mass, amard2019first}.
This may reflect a decrease in the AM-loss rate of Sun-like stars at the age of the Sun, as proposed by \cite{van2016weakened}, though this value perhaps indicates a less abrupt change to the AM-loss rate.
On the other hand, models of stellar rotation-evolution (relying on measured rotation rates of stars at various ages) currently only probe the AM-loss as averaged over timescales of $\sim 10-100$Myrs.
Historical estimates of the solar AM-loss rate are currently limited by the available reconstructions of solar activity, which are confined by the last ice-age \citep[see][]{finley2019solar}. Over this period of around 9000yrs, it is possible that the Sun had a reduced magnetic activity compared to other Sun-like stars \citep[e.g.][]{reinhold2020sun}, which should also be reflected in a weaker AM-loss rate.
Thus, if the Sun's magnetism varies on much longer timescales than can currently be measured, AM-loss rates recovered from spacecraft observations would remain ambiguous in the context of stellar spin-evolution.
However, this remains an interesting connection between the Sun and other Sun-like stars that will continue to be investigated using concurrent multi-spacecraft observations of the solar wind AM flux (at various radial distances), facilitated by PSP \citep{fox2016solar}, the \textit{Solar Orbiter} spacecraft \citep{mueller2013solar}, and existing instruments at 1au. Additionally, with solar activity increasing as the Sun enters solar cycle 25, such multi-spacecraft observations will be able to study the influence of varying activity on the solar AM-loss rate.


\acknowledgments
We thank the SWEAP and FIELDS instrument teams of Parker Solar Probe, and the NASA/GSFC's Space Physics Data Facility for providing this data.
Parker Solar Probe was designed, built, and is now operated by the Johns Hopkins Applied Physics Laboratory as part of NASA’s Living with a Star (LWS) program (contract NNN06AA01C). Support from the LWS management and technical team has played a critical role in the success of the Parker Solar Probe mission.
AJF and SPM acknowledge funding from the European Research Council (ERC) under the European Union’s Horizon 2020 research and innovation programme (grant agreement No 682393 AWESoMeStars).
VR acknowledges funding by the ERC SLOW{\_}\,SOURCE project (SLOW{\_}\,SOURCE - DLV-819189).
MO is funded by Science and Technology Facilities Council (STFC) grant numbers ST/M000885/1
and ST/R000921/1.
RFP acknowledges support from the French space agency (Centre National des Etudes 624 Spatiales; CNES; https://cnes.fr/fr).
Figures in this work are produced using the python package matplotlib \citep{hunter2007matplotlib}.

\appendix

\section{Third Encounter Data}\label{E03}
\begin{figure}
   \begin{center}
    \includegraphics[trim=0.cm 1.cm 0.cm 0.cm, clip, width=0.7\textwidth]{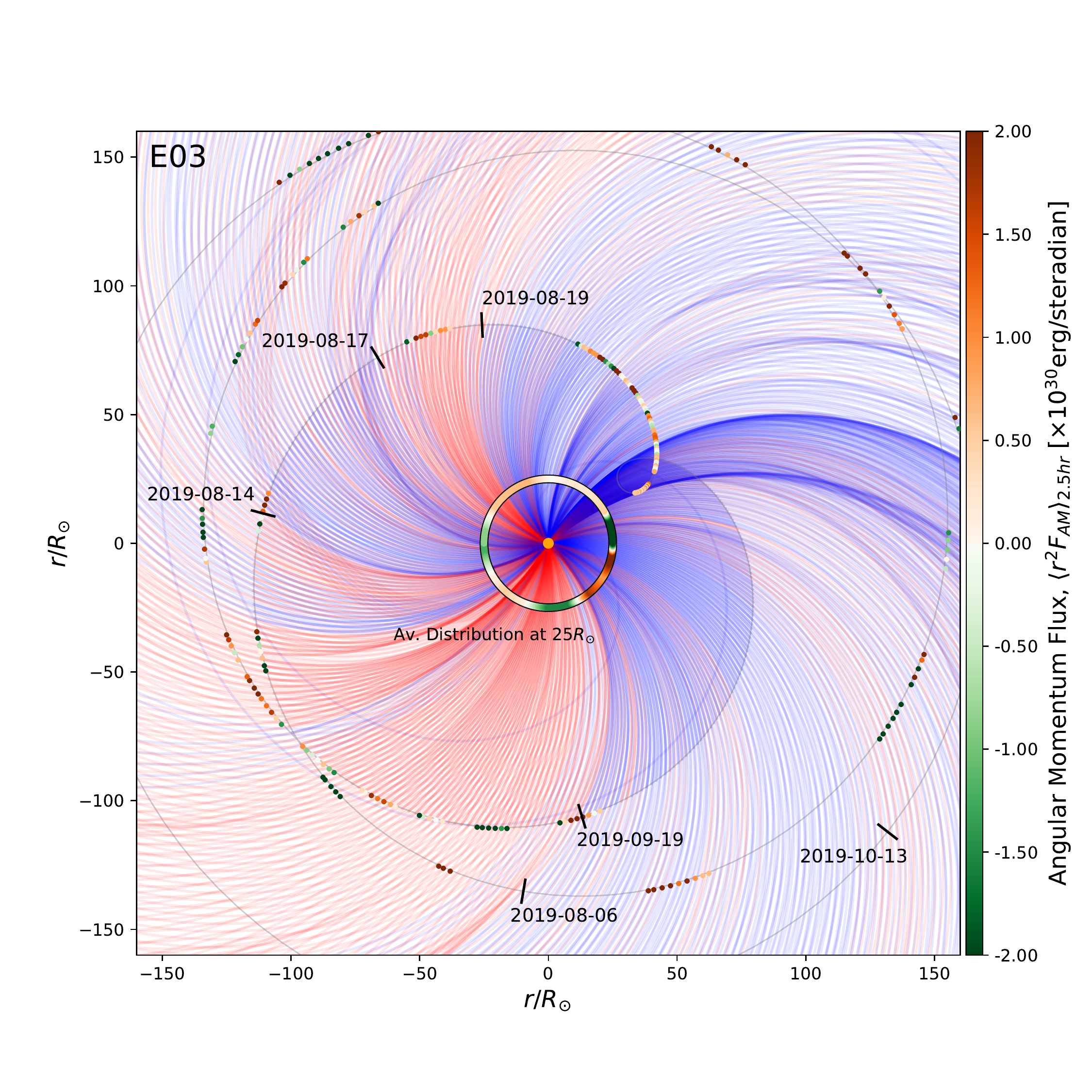}\\
    \includegraphics[trim=0.cm 1.cm 0.cm 0.cm, clip, width=\textwidth]{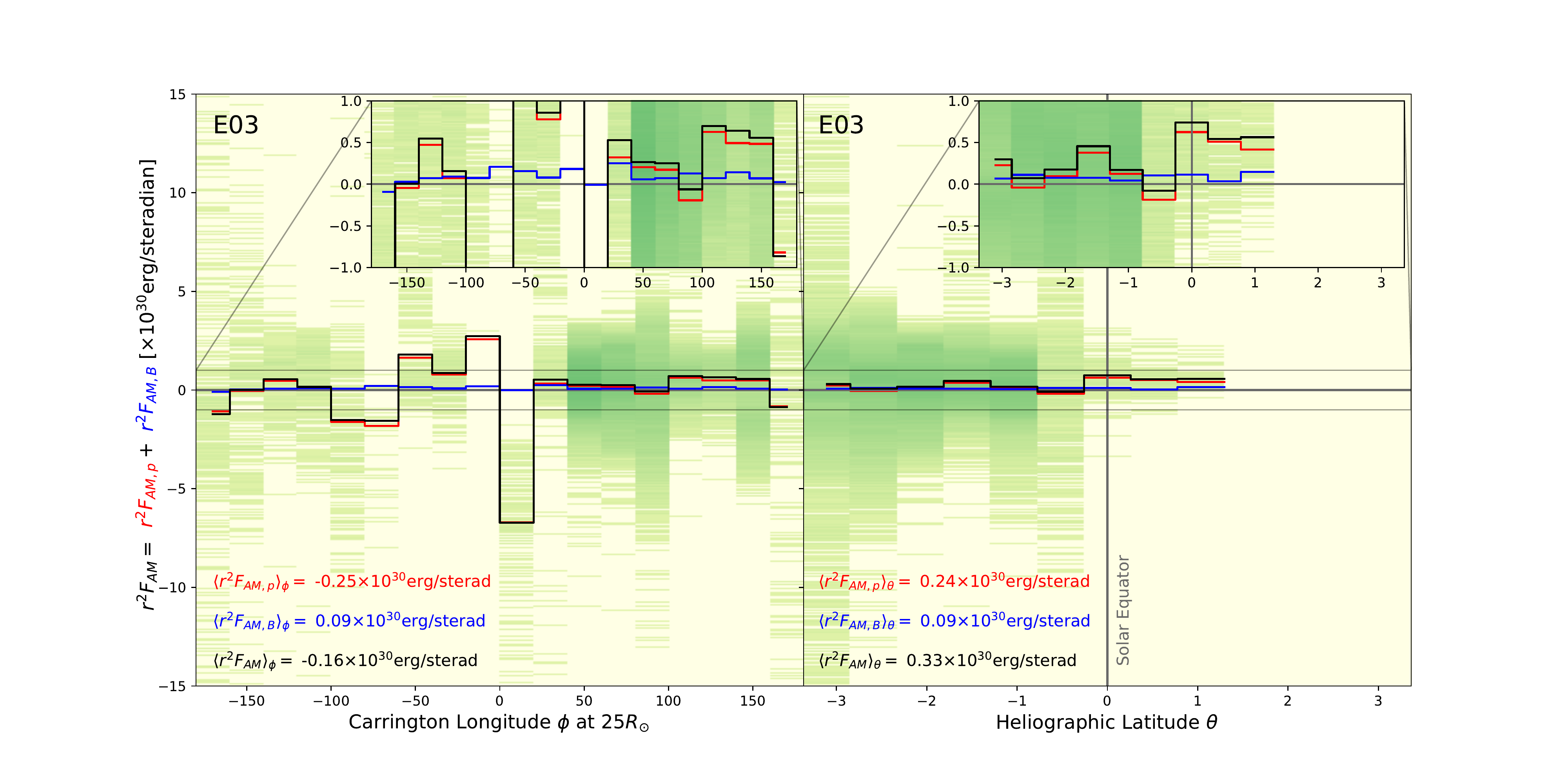}
  \end{center}
   \caption{Top: Similar to Figure \ref{trajectory}. AM flux as observed by PSP during its third orbit, shown in a rotating frame with the Sun. Bottom: Similar to Figures \ref{longitudinallyAveraged} and \ref{latitudinallyAveraged}, now for the third orbit of PSP.}
   \label{e03data}
\end{figure}

During PSP's third orbit, the SPC instrument stopped taking data during its approach to perihelion, which significantly reduced the data available from this orbit. Using the publicly available dataset, we perform the same analysis as in Section 2. The results from this are shown in Figure \ref{e03data}. As with E01 and E02, the magnetic stresses are consistently around $0.1\times 10^{30}$erg whereas the AM flux in the protons shows strong positive/negative variations with longitude. It is however difficult to compare this with the previous two orbits given the stark differences in spatial/temporal sampling. Surprisingly, the average AM flux (when the data are binned versus heliographic latitude) is similar to that gained by analysing E01 and E02, though this is likely coincidental.

\bibliographystyle{yahapj}
\bibliography{papers}

\begin{thebibliography}{}
\providecommand\natexlab[1]{#1}
\providecommand\JournalTitle[1]{#1}

\bibitem[{Ag{\"u}eros(2017)}]{agueros2017setting}
Ag{\"u}eros, M. 2017, \JournalTitle{Revista Mexicana de Astronom{\'i}a y
  Astrof{\'i}sica}, 49, 80

\bibitem[{Ag{\"u}eros {et~al.}(2011)Ag{\"u}eros, Covey, Lemonias, Law, Kraus,
  Batalha, Bloom, Cenko, Kasliwal, Kulkarni, {et~al.}}]{agueros2011factory}
Ag{\"u}eros, M., Covey, K., Lemonias, J., {et~al.} 2011, \JournalTitle{The
  Astrophysical Journal}, 740, 110

\bibitem[{Amard {et~al.}(2016)Amard, Palacios, Charbonnel, Gallet, \&
  Bouvier}]{amard2016rotating}
Amard, L., Palacios, A., Charbonnel, C., Gallet, F., \& Bouvier, J. 2016,
  \JournalTitle{Astronomy \& Astrophysics}, 587, A105

\bibitem[{Amard {et~al.}(2019)Amard, Palacios, Charbonnel, Gallet, Georgy,
  Lagarde, \& Siess}]{amard2019first}
Amard, L., Palacios, A., Charbonnel, C., {et~al.} 2019, \JournalTitle{Astronomy
  \& Astrophysics}, 631, A77

\bibitem[{Bale {et~al.}(2016)Bale, Goetz, Harvey, Turin, Bonnell, De~Wit,
  Ergun, MacDowall, Pulupa, Andr{\'e}, {et~al.}}]{bale2016fields}
Bale, S., Goetz, K., Harvey, P., {et~al.} 2016, \JournalTitle{Space science
  reviews}, 204, 49

\bibitem[{Bale {et~al.}(2019)Bale, Badman, Bonnell, Bowen, Burgess, Case,
  Cattell, Chandran, Chaston, Chen, {et~al.}}]{bale2019highly}
Bale, S., Badman, S., Bonnell, J., {et~al.} 2019, \JournalTitle{Nature}, 576,
  237

\bibitem[{Barnes(2003)}]{barnes2003rotational}
Barnes, S. 2003, \JournalTitle{The Astrophysical Journal}, 586, 464

\bibitem[{Barnes \& Kim(2010)}]{barnes2010angular}
Barnes, S., \& Kim, Y. 2010, \JournalTitle{The Astrophysical Journal}, 721, 675

\bibitem[{Blackman \& Owen(2016)}]{blackman2016minimalist}
Blackman, E., \& Owen, J. 2016, \JournalTitle{Monthly Notices of the Royal
  Astronomical Society}, 458, 1548

\bibitem[{Brown(2014)}]{brown2014metastable}
Brown, T. 2014, \JournalTitle{The Astrophysical Journal}, 789, 101

\bibitem[{Brun \& Browning(2017)}]{brun2017magnetism}
Brun, S., \& Browning, M. 2017, \JournalTitle{Living Reviews in Solar Physics},
  14, 4

\bibitem[{Case {et~al.}(2020)Case, Kasper, Stevens, Korreck, Paulson, Daigneau,
  Caldwell, Freeman, Henry, Klingensmith, {et~al.}}]{case2020solar}
Case, A.~W., Kasper, J.~C., Stevens, M.~L., {et~al.} 2020, \JournalTitle{The
  Astrophysical Journal Supplement Series}, 246, 43

\bibitem[{Covey {et~al.}(2016)Covey, Ag{\"u}eros, Law, Liu, Ahmadi, Laher,
  Levitan, Sesar, \& Surace}]{covey2016rapidly}
Covey, K., Ag{\"u}eros, M., Law, N., {et~al.} 2016, \JournalTitle{The
  Astrophysical Journal}, 822, 81

\bibitem[{Crooker {et~al.}(2010)Crooker, Appleton, Schwadron, \&
  Owens}]{crooker2010suprathermal}
Crooker, N., Appleton, E., Schwadron, N., \& Owens, M. 2010,
  \JournalTitle{Journal of Geophysical Research: Space Physics}, 115

\bibitem[{Curtis {et~al.}(2019)Curtis, Ag{\"u}eros, Douglas, \&
  Meibom}]{curtis2019temporary}
Curtis, J., Ag{\"u}eros, M., Douglas, S., \& Meibom, S. 2019, \JournalTitle{The
  Astrophysical Journal}, 879, 49

\bibitem[{Davenport \& Covey(2018)}]{davenport2018rotating}
Davenport, J., \& Covey, K. 2018, \JournalTitle{The Astrophysical Journal},
  868, 151

\bibitem[{Douglas {et~al.}(2017)Douglas, Ag{\"u}eros, Covey, \&
  K}]{douglas2017poking}
Douglas, S., Ag{\"u}eros, M., Covey, K., \& K, A. 2017, \JournalTitle{The
  Astrophysical Journal}, 842, 83

\bibitem[{Finley {et~al.}(2019{\natexlab{a}})Finley, Deshmukh, Matt, Owens, \&
  Wu}]{finley2019solar}
Finley, A., Deshmukh, S., Matt, S., Owens, M., \& Wu, C. 2019{\natexlab{a}},
  \JournalTitle{The Astrophysical Journal}, 883, 67

\bibitem[{Finley {et~al.}(2018)Finley, Matt, \& See}]{finley2018effect}
Finley, A., Matt, S., \& See, V. 2018, \JournalTitle{The Astrophysical
  Journal}, 864, 125

\bibitem[{Finley {et~al.}(2019{\natexlab{b}})Finley, Hewitt, Matt, Owens,
  Pinto, \& R{\'e}ville}]{finley2019direct}
Finley, A.~J., Hewitt, A.~L., Matt, S.~P., {et~al.} 2019{\natexlab{b}},
  \JournalTitle{The Astrophysical Journal Letters}, 885, L30

\bibitem[{Fisk \& Kasper(2020)}]{fisk2020global}
Fisk, L., \& Kasper, J. 2020, \JournalTitle{The Astrophysical Journal Letters},
  894, L4

\bibitem[{Fox {et~al.}(2016)Fox, Velli, Bale, Decker, Driesman, Howard, Kasper,
  Kinnison, Kusterer, Lario, {et~al.}}]{fox2016solar}
Fox, N., Velli, M., Bale, S., {et~al.} 2016, \JournalTitle{Space Science
  Reviews}, 204, 7

\bibitem[{Gallet \& Bouvier(2013)}]{gallet2013improved}
Gallet, F., \& Bouvier, J. 2013, \JournalTitle{Astronomy \& Astrophysics}, 556,
  A36

\bibitem[{Gallet \& Bouvier(2015)}]{gallet2015improved}
---. 2015, \JournalTitle{Astronomy \& Astrophysics}, 577, A98

\bibitem[{Gallet {et~al.}(2017)Gallet, Charbonnel, Amard, Brun, Palacios, \&
  Mathis}]{gallet2017impacts}
Gallet, F., Charbonnel, C., Amard, L., {et~al.} 2017, \JournalTitle{Astronomy
  \& Astrophysics}, 597, A14

\bibitem[{Garraffo {et~al.}(2018)Garraffo, Drake, Dotter, Choi, Burke, Moschou,
  Alvarado-G{\'o}mez, Kashyap, \& Cohen}]{garraffo2018revolution}
Garraffo, C., Drake, J., Dotter, A., {et~al.} 2018, \JournalTitle{The
  Astrophysical Journal}, 862, 90

\bibitem[{Goedbloed {et~al.}(2019)Goedbloed, Keppens, \&
  Poedts}]{goedbloed2019magnetohydrodynamics}
Goedbloed, H., Keppens, R., \& Poedts, S. 2019, Magnetohydrodynamics: Of
  Laboratory and Astrophysical Plasmas (Cambridge University Press)

\bibitem[{Hunter(2007)}]{hunter2007matplotlib}
Hunter, J. 2007, \JournalTitle{Computing In Science \& Engineering}, 9, 90

\bibitem[{Johnstone {et~al.}(2015{\natexlab{a}})Johnstone, G{\"u}del,
  L{\"u}ftinger, Toth, \& Brott}]{johnstone2015stellar}
Johnstone, C., G{\"u}del, M., L{\"u}ftinger, T., Toth, G., \& Brott, I.
  2015{\natexlab{a}}, \JournalTitle{Astronomy \& Astrophysics}, 577, A27

\bibitem[{Johnstone {et~al.}(2015{\natexlab{b}})Johnstone, G{\"u}del,
  St{\"o}kl, Lammer, Tu, Kislyakova, L{\"u}ftinger, Odert, Erkaev, \&
  Dorfi}]{johnstone2015evolution}
Johnstone, C., G{\"u}del, M., St{\"o}kl, A., {et~al.} 2015{\natexlab{b}},
  \JournalTitle{The Astrophysical Journal Letters}, 815, L12

\bibitem[{Kasper {et~al.}(2019)Kasper, Bale, Belcher, Berthomier, Case,
  Chandran, Curtis, Gallagher, Gary, Golub, {et~al.}}]{kasper2019alfvenic}
Kasper, J., Bale, S., Belcher, J., {et~al.} 2019, \JournalTitle{Nature}, 576,
  228

\bibitem[{Kasper {et~al.}(2016)Kasper, Abiad, Austin, Balat-Pichelin, Bale,
  Belcher, Berg, Bergner, Berthomier, Bookbinder, {et~al.}}]{kasper2016solar}
Kasper, J.~C., Abiad, R., Austin, G., {et~al.} 2016, \JournalTitle{Space
  Science Reviews}, 204, 131

\bibitem[{Lazarus \& Goldstein(1971)}]{lazarus1971observation}
Lazarus, A., \& Goldstein, B. 1971, \JournalTitle{The Astrophysical Journal},
  168, 571

\bibitem[{Li(1999)}]{li1999magnetic}
Li, J. 1999, \JournalTitle{Monthly Notices of the Royal Astronomical Society},
  302, 203

\bibitem[{Lorenzo-Oliveira {et~al.}(2019)Lorenzo-Oliveira, Mel{\'e}ndez,
  Yana~Galarza, Ponte, dos Santos, Spina, Bedell, Ram{\'\i}rez, Bean, \&
  Asplund}]{lorenzo2019constraining}
Lorenzo-Oliveira, D., Mel{\'e}ndez, J., Yana~Galarza, J., {et~al.} 2019,
  \JournalTitle{Monthly Notices of the Royal Astronomical Society: Letters},
  485, L68

\bibitem[{Marsch(1986)}]{marsch1986acceleration}
Marsch, E. 1986, \JournalTitle{Astronomy and Astrophysics}, 164, 77

\bibitem[{Marsch \& Richter(1984{\natexlab{a}})}]{marsch1984distribution}
Marsch, E., \& Richter, A. 1984{\natexlab{a}}, \JournalTitle{Journal of
  Geophysical Research: Space Physics}, 89, 5386

\bibitem[{Marsch \& Richter(1984{\natexlab{b}})}]{marsch1984helios}
---. 1984{\natexlab{b}}, \JournalTitle{Journal of Geophysical Research: Space
  Physics}, 89, 6599

\bibitem[{Matt {et~al.}(2015)Matt, Brun, Baraffe, Bouvier, \&
  Chabrier}]{matt2015mass}
Matt, S., Brun, S., Baraffe, I., Bouvier, J., \& Chabrier, G. 2015,
  \JournalTitle{The Astrophysical Journal Letters}, 799, L23

\bibitem[{McManus {et~al.}(2020)McManus, Bowen, Mallet, Chen, Chandran, Bale,
  Larson, de~Wit, Kasper, Stevens, {et~al.}}]{mcmanus2020cross}
McManus, M.~D., Bowen, T.~A., Mallet, A., {et~al.} 2020, \JournalTitle{The
  Astrophysical Journal Supplement Series}, 246, 67

\bibitem[{McQuillan {et~al.}(2013)McQuillan, Aigrain, \&
  Mazeh}]{mcquillan2013measuring}
McQuillan, A., Aigrain, S., \& Mazeh, T. 2013, \JournalTitle{Monthly Notices of
  the Royal Astronomical Society}, 432, 1203

\bibitem[{Mestel(1968)}]{mestel1968magnetic}
Mestel, L. 1968, \JournalTitle{Monthly Notices of the Royal Astronomical
  Society}, 138, 359

\bibitem[{Metcalfe \& Egeland(2019)}]{metcalfe2019understanding}
Metcalfe, T., \& Egeland, R. 2019, \JournalTitle{The Astrophysical Journal},
  871, 39

\bibitem[{Mozer {et~al.}(2020)Mozer, Agapitov, Bale, Bonnell, Case, Chaston,
  Curtis, de~Wit, Goetz, Goodrich, {et~al.}}]{mozer2020switchbacks}
Mozer, F., Agapitov, O., Bale, S., {et~al.} 2020, \JournalTitle{The
  Astrophysical Journal Supplement Series}, 246, 68

\bibitem[{Mueller {et~al.}(2013)Mueller, Marsden, Cyr, Gilbert,
  {et~al.}}]{mueller2013solar}
Mueller, D., Marsden, R., Cyr, O.~S., Gilbert, H., {et~al.} 2013,
  \JournalTitle{Solar Physics}, 285, 25

\bibitem[{Nascimento~Jr {et~al.}(2020)Nascimento~Jr, de~Almeida, Velloso,
  Anthony, Barnes, Saar, Meibom, da~Costa, Castro, Galarza,
  {et~al.}}]{nascimento2020rotation}
Nascimento~Jr, J.-D.~d., de~Almeida, L., Velloso, E.~N., {et~al.} 2020,
  \JournalTitle{arXiv preprint arXiv:2006.06204}

\bibitem[{N{\'u}{\~n}ez {et~al.}(2015)N{\'u}{\~n}ez, Ag{\"u}eros, Covey,
  Hartman, Kraus, Bowsher, Douglas, L{\'o}pez-Morales, Pooley, Posselt,
  {et~al.}}]{nunez2015linking}
N{\'u}{\~n}ez, A., Ag{\"u}eros, M., Covey, K., {et~al.} 2015, \JournalTitle{The
  Astrophysical Journal}, 809, 161

\bibitem[{{Panasenco} {et~al.}(2020){Panasenco}, {Velli}, {D'Amicis}, {Shi},
  {R{\'e}ville}, {Bale}, {Badman}, {Kasper}, {Korreck}, {Bonnell}, {Wit},
  {Goetz}, {Harvey}, {MacDowall}, {Malaspina}, {Pulupa}, {Case}, {Larson},
  {Livi}, {Stevens}, \& {Whittlesey}}]{Panasenco2020ApJS}
{Panasenco}, O., {Velli}, M., {D'Amicis}, R., {et~al.} 2020,
  \href{http://dx.doi.org/10.3847/1538-4365/ab61f4}{\JournalTitle{\apjs}, 246,
  54}

\bibitem[{Pizzo(1978)}]{pizzo1978three}
Pizzo, V. 1978, \JournalTitle{Journal of Geophysical Research: Space Physics},
  83, 5563

\bibitem[{Pizzo {et~al.}(1983)Pizzo, Schwenn, Marsch, Rosenbauer,
  M{\"u}hlh{\"a}user, \& Neubauer}]{pizzo1983determination}
Pizzo, V., Schwenn, R., Marsch, E., {et~al.} 1983, \JournalTitle{The
  Astrophysical Journal}, 271, 335

\bibitem[{Rebull {et~al.}(2016)Rebull, Stauffer, Bouvier, Cody, Hillenbrand,
  Soderblom, Valenti, Barrado, Bouy, Ciardi, {et~al.}}]{rebull2016rotation}
Rebull, L., Stauffer, J., Bouvier, J., {et~al.} 2016, \JournalTitle{The
  Astronomical Journal}, 152, 113

\bibitem[{Reinhold {et~al.}(2019)Reinhold, Bell, Kuszlewicz, Hekker, \&
  Shapiro}]{reinhold2019transition}
Reinhold, T., Bell, K., Kuszlewicz, J., Hekker, S., \& Shapiro, A. 2019,
  \JournalTitle{Astronomy \& Astrophysics}, 621, A21

\bibitem[{Reinhold {et~al.}(2020)Reinhold, Shapiro, Solanki, Montet, Krivova,
  Cameron, \& Amazo-G{\'o}mez}]{reinhold2020sun}
Reinhold, T., Shapiro, A.~I., Solanki, S.~K., {et~al.} 2020,
  \JournalTitle{Science}, 368, 518

\bibitem[{R{\'e}ville {et~al.}(2020{\natexlab{a}})R{\'e}ville, Velli,
  Rouillard, Lavraud, Tenerani, Shi, \& Strugarek}]{reville2020tearing}
R{\'e}ville, V., Velli, M., Rouillard, A.~P., {et~al.} 2020{\natexlab{a}},
  \JournalTitle{The Astrophysical Journal Letters}, 895, L20

\bibitem[{R{\'e}ville {et~al.}(2020{\natexlab{b}})R{\'e}ville, Velli,
  Panasenco, Tenerani, Shi, Badman, Bale, Kasper, Stevens, Korreck,
  {et~al.}}]{reville2020role}
R{\'e}ville, V., Velli, M., Panasenco, O., {et~al.} 2020{\natexlab{b}},
  \JournalTitle{The Astrophysical Journal Supplement Series}, 246, 24

\bibitem[{Rouillard {et~al.}(2020)Rouillard, Kouloumvakos, Vourlidas, Kasper,
  Bale, Raouafi, Lavraud, Howard, Stenborg, Stevens,
  {et~al.}}]{rouillard2020relating}
Rouillard, A.~P., Kouloumvakos, A., Vourlidas, A., {et~al.} 2020,
  \JournalTitle{The Astrophysical Journal Supplement Series}, 246, 37

\bibitem[{Sadeghi~Ardestani {et~al.}(2017)Sadeghi~Ardestani, Guillot, \&
  Morel}]{sadeghi2017semi}
Sadeghi~Ardestani, L., Guillot, T., \& Morel, P. 2017, \JournalTitle{Monthly
  Notices of the Royal Astronomical Society}, 472, 2590

\bibitem[{See {et~al.}(2018)See, Jardine, Vidotto, Donati, Saikia, Fares,
  Folsom, Jeffers, Marsden, Morin, {et~al.}}]{see2018open}
See, V., Jardine, M., Vidotto, A., {et~al.} 2018, \JournalTitle{Monthly Notices
  of the Royal Astronomical Society}, 474, 536

\bibitem[{Skumanich(1972)}]{skumanich1972time}
Skumanich, A. 1972, \JournalTitle{The Astrophysical Journal}, 171, 565

\bibitem[{Soderblom(1983)}]{soderblom1983rotational}
Soderblom, D. 1983, \JournalTitle{The Astrophysical Journal Supplement Series},
  53, 1

\bibitem[{van Saders {et~al.}(2016)van Saders, Ceillier, Metcalfe, Aguirre,
  Pinsonneault, Garc{\'\i}a, Mathur, \& Davies}]{van2016weakened}
van Saders, J., Ceillier, T., Metcalfe, T., {et~al.} 2016,
  \JournalTitle{Nature}, 529, 181

\bibitem[{Weber \& Davis(1967)}]{weber1967angular}
Weber, E., \& Davis, L. 1967, \JournalTitle{The Astrophysical Journal}, 148,
  217

\bibitem[{Whittlesey {et~al.}(2020)Whittlesey, Larson, Kasper, Halekas,
  Abatcha, Abiad, Berthomier, Case, Chen, Curtis,
  {et~al.}}]{whittlesey2020solar}
Whittlesey, P.~L., Larson, D.~E., Kasper, J.~C., {et~al.} 2020,
  \JournalTitle{The Astrophysical Journal Supplement Series}, 246, 74

\bibitem[{Wright \& Drake(2016)}]{wright2016solar}
Wright, N., \& Drake, J. 2016, \JournalTitle{Nature}, 535, 526

\bibitem[{Wright {et~al.}(2011)Wright, Drake, Mamajek, \&
  Henry}]{wright2011stellar}
Wright, N., Drake, J., Mamajek, E., \& Henry, G. 2011, \JournalTitle{The
  Astrophysical Journal}, 743, 48

\end{thebibliography}



\end{document}